\documentclass{article}

\usepackage[T1]{fontenc}
\usepackage{amssymb}
\usepackage{url}  
\usepackage{subfigure}
\usepackage{amsmath, amssymb}
\usepackage{amsfonts,amsmath,amssymb,epsfig,epsf,psfrag}
\usepackage[english]{babel}
\usepackage{enumerate}
\usepackage{url}
\usepackage{color}
\usepackage[utf8]{inputenc}
\usepackage{lmodern}
\usepackage{graphicx}
\usepackage[T1]{fontenc}
\usepackage{amsthm}
\usepackage{multimedia}
\usepackage{changebar}
\usepackage{natbib}
\newcommand{\ICL}{\mbox{ICL}}

\begin{document}

\title{Fast estimation of the Integrated Completed Likelihood criterion for change-point detection problems with applications to Next-Generation Sequencing data}
\author{A. Cleynen$^{1}$; The Minh Luong$^{2}$; Guillem Rigaill$^{3}$ and Gregory Nuel$^{2}$}
\date{}
\maketitle \noindent $^{1}${\it AgroParisTech, UMR 518 MIA, 16, rue Claude Bernard, 75005 Paris, France\\
INRA, UMR 518 MIA, 16, rue Claude Bernard, 75005 Paris, France.} \\
\noindent $^{2}${\it MAP5 - UMR CNRS 8145, Universit\'e Paris Descartes, Paris, France} \\
\noindent $^{3}${\it URGV INRA-CNRS-Universit\'e d'\'Evry Val d'Essonne, 2 Rue Gaston Cr\'emieux,\\ 91057 Evry Cedex, France} \\

\begin{abstract}
In this paper, we consider the Integrated Completed Likelihood (ICL) as a useful criterion for estimating the number of changes in the underlying distribution of data, specifically in problems where detecting the precise location of these changes is the main goal.
The exact computation of the ICL requires $\mathcal{O}(Kn^2)$ operations (with $K$ the number of segments and $n$ the number of data-points) which is prohibitive in many practical situations with large sequences of data.
We describe a framework to estimate the ICL with $\mathcal{O}(Kn)$ complexity.
Our approach is general in the sense that it can accommodate any given model distribution.
We checked the run-time and validity of our approach on simulated data and demonstrate its good performance when analyzing real Next-Generation Sequencing (NGS) data using a negative binomial model. Our method is implemented in the R package \texttt{postCP} and available on the CRAN repository.
\end{abstract}

Hidden Markov Model ; Integrated Completed Likelihood ; Model Selection ; Negative Binomial ; Segmentation

\section{Introduction}
\label{Introduction}

The estimation of the number of segments is a central aspect in change-point methodology. For instance, in the context of CGH-array or Next-Generation Sequencing experiments, identifying the number and corresponding location of segments is crucial as the segments may relate to a biological event of interest. This theoretically complex problem can be handled in the more general context of model selection, leading to the use of \textit{ad hoc} procedures in practical situations.

Among the procedures are the use of classical criteria based on penalized likelihoods such as the Akaike Information Criterion (AIC) and the Bayes Information Criterion \citep[BIC or SIC,][]{Yao88}. However, when choosing the number of segments, the BIC criterion uses a Laplace approximation requiring differentiability conditions not satisfied by the model, which thus may not be appropriate when the number of observations in each segment are unequal and unknown. These criteria also tend to overestimate the number of segments as the clustering within segments tends to be ignored, as shown by \cite{zhang_modified_2007} who proposed a modified BIC criterion using a Brownian motion model with changing drift for the specific case of normal data. 

For this reason, there has been an extensive literature influenced by \cite{birge2001gaussian} which proposes new penalty shapes and constants in order to select a lower number of segments in the profile. The idea is to choose the model that, within a set of models, performs closest to the true value by deriving a tight upper bound on the variance term. This leads to penalties that generally depend only on the number of segments $K$, and whose constants can be chosen adaptively to the data \citep{lavielle_using_2005,lebarbier_detecting_2005}. However, a large proportion of those methods focused on normal data, and are not applicable to count datasets modeled by the Poisson or the negative binomial distributions.
 
Other approaches for model selection appearing in the literature include sequential likelihood ratio tests \citep{doi:10.1080/00949658808811102} and Bayesian approaches through estimating model posterior probabilities by various MCMC methods  \citep{green1995reversible, chib1998estimation, andrieu2001model, fearnhead_exact_2005}. However, the Bayesian approaches are often computationally intensive as they require re-sampling.

In the context of incomplete data models (e.g. mixture model for clustering) \cite{biernacki_assessingmixture_2000} proposed a model selection criterion accounting for both observed and unobserved variables based on the Integrated Completed Likelihood (ICL): $\sum_S \mathbb{P}(S|X) \log \mathbb{P}(S|X)$ where $X$ are the observations and $S$ are the corresponding (unknown) clustering membership.

\cite{rigaill_exact_2011} proposed the use of the ICL criterion in the multiple change-point detection context. Hence, the segmentation $S$ can be considered as {a set of unobserved variables} in the sense that the segment-labels of each datapoint are not known. In this context, we can select the number of segments as:
  \begin{equation}
  \hat{K} = \arg\min_{K} \ICL(K) \quad \text{where} \quad \ICL(K) = - \log \mathbb{P}(X , K) + \mathcal{H}(K),\label{ICL}
  \end{equation} 
with  $\mathcal{H}(K) = - \sum_{S \in \mathcal{M}_K} \mathbb{P}(S | X, K) \log \mathbb{P}(S | X, K)$, and $\mathcal{M}_K$ representing the set of all segmentations of the signal in $K$ segments.

The entropy term $\mathcal{H}(K)$ can be viewed as an intrinsic penalty to quantify the reliability of a given model with $K$ segments by characterizing the separation of the observations in different segments.  In other words, for fixed $K$ segments, the entropy $\mathcal{H}(K)$ thus will be lower when the best segmentation provides a much better fit compared to other segmentations with the same number of segments, hence favoring models which provide the most evidence of {similarity} within the detected segments.
While other penalized likelihood approaches are designed to select the most likely number of segments by relying on approximation of posterior probabilities or oracle inequalities, the ICL criterion aims at selecting the number of segments with the lowest uncertainty.

In the context of Hidden Markov Models (HMMs), it is well known that the posterior distribution $\mathbb{P}(S | X, K,\Theta_K)$ can be efficiently computed using standard forward-backward recursions with $\mathcal{O}(K^2n)$ complexity \citep{martin2012distribution}. However, the HMM requires that emission parameters take their values in a limited set of levels which are recurrently visited by the underlying hidden process.

In the segmentation context, where each segment has its own specific level, an exact algorithm with $\mathcal{O}(Kn^2)$ complexity computes the ICL in a Bayesian framework. In a simulation study, \cite{rigaill_exact_2011} showed that the ICL performed better than standard model selection criteria such as BIC or Deviance Information Criterion (DIC). However the quadratic complexity and numerical precision restrict the use of this Bayesian ICL to relatively small profiles. 

In this paper {we suggest a computation of the ICL conditionally to the segment parameters and} we {propose} a fast {two-step} procedure to compute {this conditional ICL} criterion with linear complexity in order to select the number of segments within a set of change-point data. {First, we specify a range of possible $K$ number of change-points, from one to a user-defined $K_{\max}$. We estimate the parameters of the segmentation in $K$ segments, and given these estimates, we compute the ICL for each value of $K$ in the range. Second, we select the $K$ which minimizes the ICL criterion.} {In essence, our conditional ICL explores only one aspect of the segmentation uncertainty, the position of the change-points, and ignores the uncertainty due to the segment parameters.}

Section \ref{methods} describes the ICL estimation procedure, through the use of a constrained hidden Markov model and Section \ref{simulation} validates the approach by presenting the results of different simulations for detecting the correct number of change-points. Finally, Section \ref{analysis} is a discussion of our method supported by a comparison with a few segmentation algorithms on data-sets simulated by re-sampling real RNA-Seq data, {and an illustration on the original dataset from an experiment on a chromosome from the yeast species} {from the same study}.

\section{Integrated Completed Likelihood criterion estimation using {a constrained} HMM} \label{methods}

  In this paper we use the following \emph{segment-based model} for the distribution of $X$ given a segmentation $S \in \mathcal{M}_K$: 
  \begin{equation}\label{eq:themodel}
    \mathbb{P}(X | S; \Theta_K)=\prod_{i=1}^{n} g_{\theta_{S_i}}\left(X_i\right) = \prod _{k=1}^K \prod_{i:S_i=k}  g_{\theta_k}\left(X_i\right)
  \end{equation}
  where $g_{\theta_{S_i}}(\cdot)$ is the parametric distribution (ex: normal, Poisson, negative binomial, etc.) with parameter $\theta_{S_i}$, $\Theta_K=(\theta_1,\ldots,\theta_K)$ is the global parameter, $S_i \in \{1,\ldots,K\}$ is the index of the segment at position $i$ (ex: $S_{1:5}=11222$ corresponds to a segmentation of $n=5$ points into $K=2$ segments with a change-point occurring between positions $2$ and $3$), and $\mathcal{M}_K$ is the set of all possible partitions of $S_1,\ldots,S_n$ with a fixed $K$ number of segments, such that $S_1=1$ and $S_n=K$, and $S_i-S_{i-1}\in \left\{0,1\right\}$ for all $i=2,\ldots,n$. 

  {One should note that although this model has the same emission probabilities as its HMM counterpart, the constraints on the sequence $S$ correspond \emph{exactly} to the segmentation model where each segment has its own level, and \emph{not} to any HMM where levels take their value in a recurring set.}

\subsection{Fast estimation of posterior quantities in ICL criterion}\label{hmmseg}

{Our goal is to compute the conditional ICL given by the following equation :
\begin{equation*}
  \hat{K} = \arg\min_{K} \ICL(K | \Theta_K) \quad   
  \end{equation*} 
  \begin{equation}
  \text{where} \quad \ICL(K | \Theta_K) = - \log \mathbb{P}(X , K | \Theta_K) + \mathcal{H}(K | \Theta_K).\label{ICL_cond}
  \end{equation} 
  
The objective of this conditional ICL is to reproduce the performance of the non-conditional ICL (in Equation \ref{ICL}). The conditional ICL criterion is well defined given a prior distribution on the segmentations: $\mathbb{P}(S, K)$. We will {only consider priors that can be decomposed as}:
$ \mathbb{P}(S, K) = \mathbb{P}(S|K) \mathbb{P}(K)$; {this choice is discussed in a later section. In both the conditional and non-conditional ICL, the first term, $\log \mathbb{P}(X , K)$ (and respectively $\log \mathbb{P}(X , K | \Theta_K)$), depends on both $\mathbb{P}(S|K)$ and $\mathbb{P}(K)$, however, the entropy term only depends on $\mathbb{P}(S|K)$.}}\newline

{To estimate this entropy term, we consider a specific hidden Markov model with constraints chosen specifically to correspond to a segmentation model \citep{luong} where the change-points separate segments consisting of contiguous observations with the same distribution. Introducing a prior distribution $\mathbb{P}(S, K)$ on any $S \in \mathcal{M}_K$, yields the posterior distribution of the segmentation:}
  \begin{equation}
    \mathbb{P}(S, K|X;\Theta_K)=\frac{\mathbb{P}(X|S, K;\Theta_K)\mathbb{P}(S, K)}{\sum_R \mathbb{P}(X|R, K;\Theta_K)\mathbb{P}(R, K)}.
  \end{equation}

{Considering the prior $ \mathbb{P}(S, K) = \mathbb{P}(S|K) \mathbb{P}(K)$ and fixing the value of $K$,}  let us assume that $S$ is a heterogeneous Markov chain over $\{1,2,\ldots,K,K+1\}$. We only allow for transitions of $0$ or $+1$ by constraining the chain with:  
  \begin{align*}
    &\mathbb{P}(S_1=1)=1 \\
    &\forall 2 \leqslant i \leqslant n,\quad \forall 1 \leqslant k \leqslant K,   \left\{ \begin{array}{cl}
      \mathbb{P}(S_i =k | S_{i-1}= k) & = 1-\eta_{k}(i)  \\
      \mathbb{P}(S_i =k+1 | S_{i-1}= k) & = \eta_{k}(i), \\   \end{array}\right.
  \end{align*}
  where { $\eta_{k}(i)$ is the transition probability from the $k^{th}$ segment to $k+1$ for observation $i$.}
  
  {In the general case where $K$ is not fixed, the choice of prior on $S$  is known to be a critical point. However previous methods include the use of non-informative priors {\citep{zhang_modified_2007}} when $K$ is fixed. For that reason, we focus on the uniform prior by setting $\eta_k(i)=\eta$ for all $k$ and $i$. Note that this particular case corresponds to the uniform prior $\mathbb{P}(S|K)=1/{n-1 \choose K-1}=1/|\mathcal{M}_K|$ which is used in \cite{rigaill_exact_2011}.} 

  To estimate the properties of the $K^{th}$ state we introduce a `junk' state $K+1$,  and for consistency we choose $\mathbb{P}(S_i=K+1|S_{i-1}=K+1)=1$. We then estimate the emission distribution by using the maximum likelihood estimate $g_{\hat{\theta}_k}(x_i)$, or alternatively the E-M algorithm.

  We define the forward and backward quantities as follows for observation $i$ and state $k$:
  For $1 \leqslant i \leqslant n-1$:
  \noindent 
  \begin{align*}
    F_i(k)&=\mathbb{P}(X_{1:i}=x_{1:i},S_i=k|\hat{\Theta}_k)\\
    B_i(k)&=\mathbb{P}(X_{i+1:n}=x_{i+1:n},S_n=k|S_i=k,\hat{\Theta}_k).
  \end{align*}

  We may use the following recursions to estimate the forward and backward quantities:
  \begin{align*}
    F_1(k)&=\left\{
    \begin{array}{ll}
      g_{\hat{\theta}_1}(x_1) & \text{if $k=1$}\\
      0 & \text{else}
    \end{array}
    \right.\\
  \end{align*}
  \begin{align*}
    F_{i+1}(k)&=\left[F_{i}(k)(1-\eta_k(i+1))+\mathbf{1}_{k>1} F_{i}(k-1) \eta_k(i+1) \right]g_{\hat{\theta}_k}(x_{i+1})\\
    B_{n-1}(k)&=\left\{
    \begin{array}{ll}
      \eta_K(n)g_{\hat{\theta}_k}(x_n) & \text{if $k=K-1$}\\
      (1-\eta_K(n)) g_{\hat{\theta}_k}(x_n) & \text{if $k=K$}\\
      0 & \text{else}
    \end{array}
    \right.\\
    B_{i-1}(k)&=(1-\eta_k(i))g_{\hat{\theta}_k}(x_i)B_{i}(k)+\mathbf{1}_{k<K} \eta_{k+1}(i) g_{\hat{\theta}_{k+1}}(x_i)B_{i}(k+1)
  \end{align*}
  
  These quantities can then be used to obtain the marginal distributions $\mu_i$ and the transition $\pi_i$, {being terms needed for the calculation of the entropy $\mathcal{H}(K|\hat{\Theta}_K)$} with:
  
  \begin{align}
  \mu_i(k)&=\frac{F_i(k)B_i(k)}{F_1(1)B_1(1)}\\
  \pi_i(k,k')&=
  \frac{\mathbb{P}(S_i=k'|S_{i-1}=k) g_{\hat{\theta}_{k}}(x_i)B_i(k')}{B_{i-1}(k)}.
\end{align}
where $$
\mathbb{P}(S_i=k'|S_{i-1}=k)=\left\{
\begin{array}{ll}
1-\eta & \text{if $k'=k$}\\
\eta & \text{if $k'=k+1$}\\
0 & \text{else}
\end{array}
\right.
.
$$   

{\subsubsection{Calculation of $\log \mathbb{P}(X, K | \Theta_K)$}}

{The non-conditional term $\mathbb{P} (X , K)$ can be written as $$\sum_{S \in \mathcal{M}_K} \mathbb{P}(S, K, X) = \sum_{S \in \mathcal{M}_K} \mathbb{P}(X | S, K) \mathbb{P}(S| K) \mathbb{P}(K).$$ 
In our constrained HMM approach we will therefore compute, for a given parameter $\hat{\Theta}_K$  for which the choice will be discussed later on, $\mathbb{P}(X,K|\hat{\Theta}_K)$ as $ \sum_{S \in M_K} \mathbb{P}(X | S, K, \hat{\Theta}_K) \mathbb{P}(S| K) \mathbb{P}(K)$, using the classic priors $\mathbb{P}(K) = \alpha$ and the previously discussed uniform prior $\mathbb{P}(S|K) = 1/{n-1 \choose K-1}$. The remaining term $\sum_{S \in M_K} \mathbb{P}(X | S, K, \hat{\Theta}_K)$ is then obtained directly using forward-backward recursions}. {Specifically, we obtain:
\begin{eqnarray*}
 \sum_{S \in M_K}\mathbb{P}(X,S\in \mathcal{M}_K|K,\hat{\Theta}_K)&=&F_1(1)B_1(1) \quad \text{and}\\
\mathbb{P}(S\in \mathcal{M}_K|K,\hat{\Theta}_K)&=&F^0_{1}(1)B^0_{1}(1)
 \end{eqnarray*}
where $F_i(k)$ and $B_i(k)$ are the HMM forward and backward recursions as described, and $F^0_{i}(k)$ and $B^0_{i}(k)$ are forward and backward terms obtained with the usual recursions where each emission probability is replaced by $1$. }

{The likelihood term is finally obtained as 
\begin{eqnarray}
\mathbb{P}(X,K|\hat{\Theta}_K)= \frac{\mathbb{P}(K)}{{n-1 \choose K-1}} \frac{F_1(1)B_1(1)}{F^0_{1}(1)B^0_{1}(1)}.
\end{eqnarray}}

\subsubsection{Estimation of $\mathcal{H}(K)$}

The $\ICL$ can be expressed as \citep{biernacki_assessingmixture_2000}:

\begin{equation}
{{\ICL}(K|\hat{\Theta}_K)=\mathbb{P}(X,K|\hat{\Theta}_K)+\mathcal{H}(K|\hat{\Theta}_K)},
\end{equation}
with the entropy term $\mathcal{H}(K)$ estimated by 
$\mathcal{H}(K|\hat{\Theta}_K)=-\sum_S \mathbb{P}(S|X,K,\hat{\Theta}_K) \log \mathbb{P}(S|X,K,\hat{\Theta}_K)$,   and $K$ being the number of segments. 

{For a fixed $K$, the constrained HMM {is an efficient way to estimate} the posterior segmentation distribution $\mathbb{P}(S | X, K,\hat{\Theta}_K)$ for a given set of parameters $\hat{\Theta}_K$. This model consists of a heterogeneous Markov chain (HMC) with marginal distribution $\mu_i(S_i)=\mathbb{P}(S_i|X,K,\hat{\Theta}_K)$ and heterogeneous transition $\pi_i(S_{i-1},S_i)= \mathbb{P}(S_i|S_{i-1},X,K,\hat{\Theta}_K)$. Those quantities can be computed with the recursive formulas as described above.}

It is hence easy \citep{hernando2005efficient} to derive the following expression for the entropy term:
  \begin{eqnarray} \label{Ent}
  \mathcal{H}(K|\hat{\Theta}_K)= &&-\left[\sum_{S_1} \mu_1(S_1) \log \mu_1(S_1) \right. \nonumber \\
  &&+ \left. \sum_{i=2}^n \sum_{S_{i-1},S_{i}} \mu_{i-1}(S_{i-1})\pi_i(S_{i-1},S_i) \log \pi_i(S_{i-1},S_i)\right]
  \end{eqnarray} 
{Note that information theory ensures that we have $0 \leqslant \mathcal{H}(K|\hat{\Theta}_K) \leqslant \log {n-1 \choose K-1}$.}

The original entropy term, $\mathcal{H}(K)$ has an expression including posterior probabilities, thus requiring the estimation of the posterior distribution of $S$ as detailed in Section \ref{hmmseg}. While it can be computed with quadratic complexity $O(Kn^2)$ \citep{rigaill_exact_2011} and intensive operations on probability matrices, its exact computation is usually intractable for large datasets of tens of thousands points or more. The forward-backward recursions of the HMM and Equation (\ref{Ent}) allow its estimation with linear complexity $O(Kn)$. {One should note that the key point for fast computation lies in the fact that we work conditionally to $\hat{\Theta}_K$ rather than considering the whole parameter space.}

\subsection{Model selection procedure using ICL}       

For any given $K$, using our constrained HMM method requires a set of initial parameters $\Theta_K = \{\hat{\theta}_k\}_{1\leq k\leq K}$. Because the quality of the results depends on the choice of those initial values, {we propose the use an effective segmentation algorithm to obtain a set of $K-1$ change-points, which can in turn be used to estimate the parameters $\Theta_k$ through maximum likelihood estimation}. 

We considered several options for the initialization algorithm: for normally distributed data we considered a K-means algorithm \citep{Hartigan/Wong:79,comte04}, which is a greedy method that minimizes the least-squares criterion, as well as binary segmentation \citep{binary_segmentation}, a fast heuristic to optimize the log-likelihood criterion. We also used the pruned dynamic programming algorithm \citep{Pdpa}, a fast algorithm to compute the optimal segmentation according to loss functions including Poisson, negative binomial or normal losses. {We then use the Viterbi algorithm \citep{viterbi1967error} to obtain the \textit{a posteriori} most probable set of change-points}.

To estimate the $\ICL$ of a change-point model with $K$ segments, we compute the posterior probabilities of interest through the forward-backward algorithm as previously described, which is implemented in the postCP package (available on the CRAN :  \url{http://cran.r-project.org/web/packages/postCP}).

This procedure is repeated for a range of possible values of $K$: $K_{\mbox{range}}=\left\{1,\ldots,K_{\max}\right\}$. We finally choose the  number of segments by minimizing the ICL criterion upon all values of $K$ in $K_{\mbox{range}}$, \textit{i.e.}
  \begin{eqnarray}\label{HMM-ICL}
    \hat{K}_{\ICL}=\underset{K\in K_{\mbox{range}}}{\arg \min }\ {\ICL(K|\hat{\Theta}_K)}. 
  \end{eqnarray}

\section{Validation}\label{simulation}

    {To validate the quality of our approach we first evaluated the impact of the initialization parameters. We implemented the Baum-Welch algorithm \citep{baum1970maximization} for use as a reference, and computed the Rand-Index between the segmentation resulting from the Baum-Welch approach to those resulting from our algorithm with different other initialization methods. The Rand-Index compares the adequacy between different segmentations by computing the proportion of concordant pairs of data-points, including the proportion of pairs that either belong to the same segment in the two competitive segmentations, or that are in different segments in both segmentations. 
 In a second step, we evaluated the results of our algorithm in terms of model selection on two sets of simulations}. 
    
\subsection{Impact of initialization parameters}

Because of the long run-time of the Baum-Welch (BW) algorithm, we considered a small simulation study where the data of size $n=1,000$ is simulated from the Poisson distribution with parameter $\lambda$ subject to $9$ change-points (at locations $100, 130, 200, 475, 500, 600, 630, 800$ and $975$) and {taking the successive values $1, 4.3, 1.15, 6$ and $4.2$ repeated twice}. On each of the $1,000$ replications, we ran our constrained HMM segmentation approach considering the number of segments to be known, but with different initialization parameters: those obtained by the Baum-Welch algorithm, those obtained by the pruned dynamic programming algorithm (PDPA), those obtained with a k-means approach and those obtained by running the Binary Segmentation (BinSeg) \citep{binary_segmentation} for the Poisson distribution.

\begin{figure}[!ht]
\begin{center}
\includegraphics[width=6cm,angle=0]{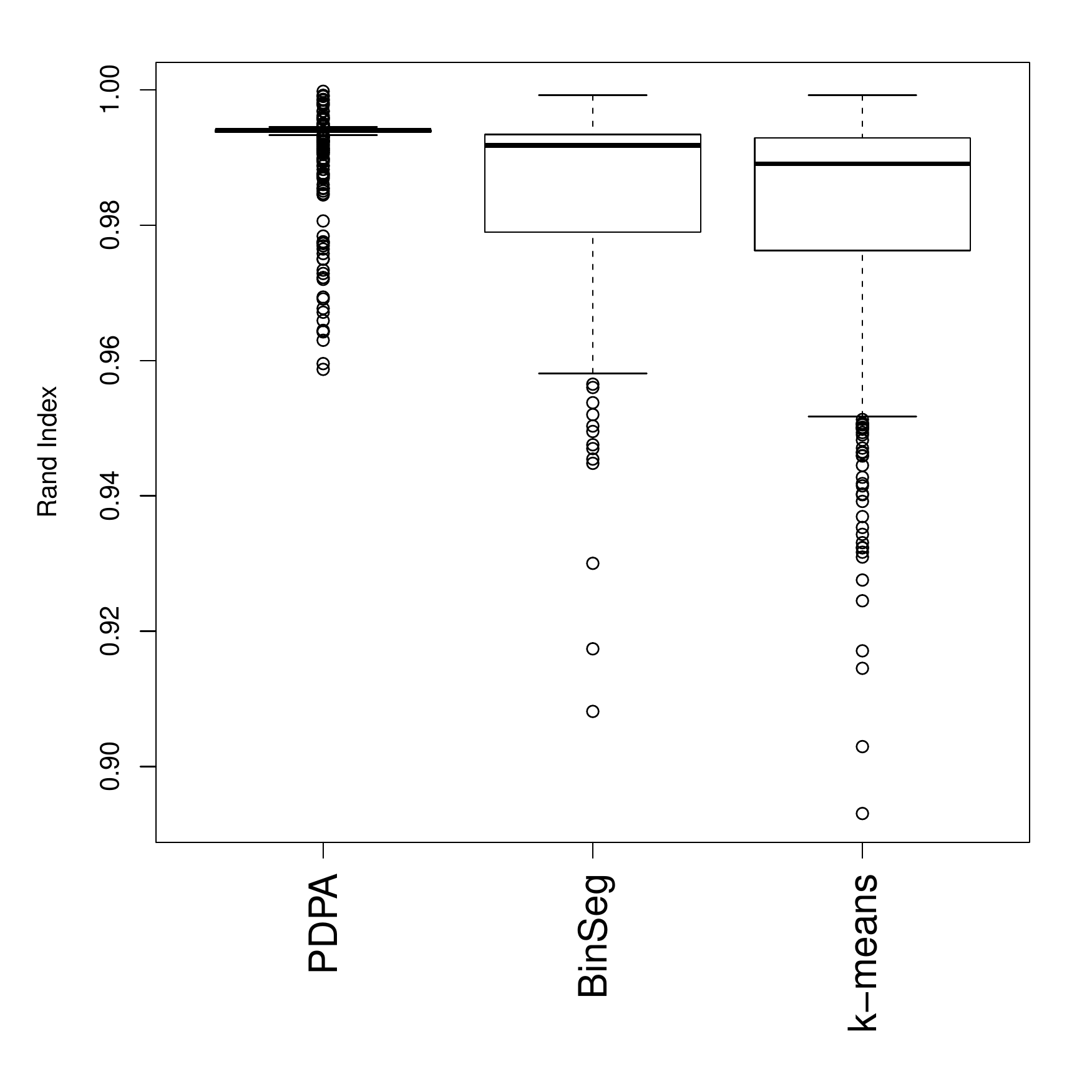} 
\end{center}
\caption{{\bf Rand-Index for the comparison of initialization methods.} Boxplot of the Rand-Index comparing the segmentation proposed by the method depending on their initialization compared to the full HMM model with Baum-Welch algorithm. As expected, the difference observed between BW and initialization with the PDPA algorithm is very small.} \label{randindex}
\end{figure}

The results are illustrated in Figure \ref{randindex}. As expected, the Rand-Index between the estimation by the Baum-Welch algorithm and the PDPA algorithm is very close to 1, and it decreases with other initialization methods that are not exact. Moreover, on average the Baum-Welch algorithm required $15.2$ iterations when itself initialized with the PDPA output, while the run-time for the initialization by PDPA requires $0.24$ seconds and an iteration of BW, $0.04$ seconds. {This finding suggests} that the combination of PDPA and postCP is advantageous in terms of run-time with a negligible difference in results, especially since the number of iterations of BW grows as $n$ and the number of segments increase (not shown here).

\subsection{Validation of the ICL approach}  
  
    Our first simulation study consisted of relatively small signals ($n=500$ points) where we compared our approach to the quadratic non-conditional algorithm. In our second simulation study, with larger signals ($n=50,000$), we only ran {our fast ICL criterion}  due to computing limitations.
  
  The simulation designs were as follows:

    \paragraph{Small design.} We used a similar simulation design suggested by \cite{rigaill_exact_2011}: we simulated a sequence of 500 observations from a Poisson model (requiring the choice of only one parameter) affected by six change-points at the following positions: $22, 65, 108, 219, 252$ and $435$. Odd segments had a mean of 1, while even segments had a mean of $1 + \lambda$, with $\lambda$ varying from $0$ to $9$. Thus, the true number of change-points were more easily identified with higher values of $\lambda$. For each configuration, we simulated {1,000} sequences.
    
    \paragraph{Large design.} We repeated the preceding procedure for large-scale datasets. We generated a sequence of {$50,000$} observations with {$K=40$} segments by randomly selecting {$39$} change-points whose locations were drawn from a uniform distribution (without replacement), with each segment needing to be at least of length $25$. For this sample size, we focus on the results from {our approximated ICL} as the non-conditional ICL implementation is not fast enough to be practical in this situation. For each configuration, we simulated {100} sequences. \newline

    We compared the performances of three different criteria:
      \begin{itemize}
        \item The {conditional} ICL \textit{greedy} (C-ICL-g) criterion where initial parameters are obtained by the greedy algorithm using least-squares, and using the criterion described in the previous section and given by Equation~(\ref{HMM-ICL}) .
        \item The {conditional} ICL \textit{exact} (C-ICL-e) criterion which corresponds to an initialization of the parameters using the pruned dynamic programming algorithm with Poisson loss.
        \item The {non-conditional} ICL (NC-ICL) criterion as described in \cite{rigaill_exact_2011}. The hyper-parameters used for the prior on the data-distribution were set to 1. {This choice is discussed in the previous paper. In this simple scenario, the results were robust to changes in the hyper-parameters (result not shown).}
      \end{itemize}

\begin{figure}[!ht]
\begin{center}
\includegraphics[width=8cm,angle=0]{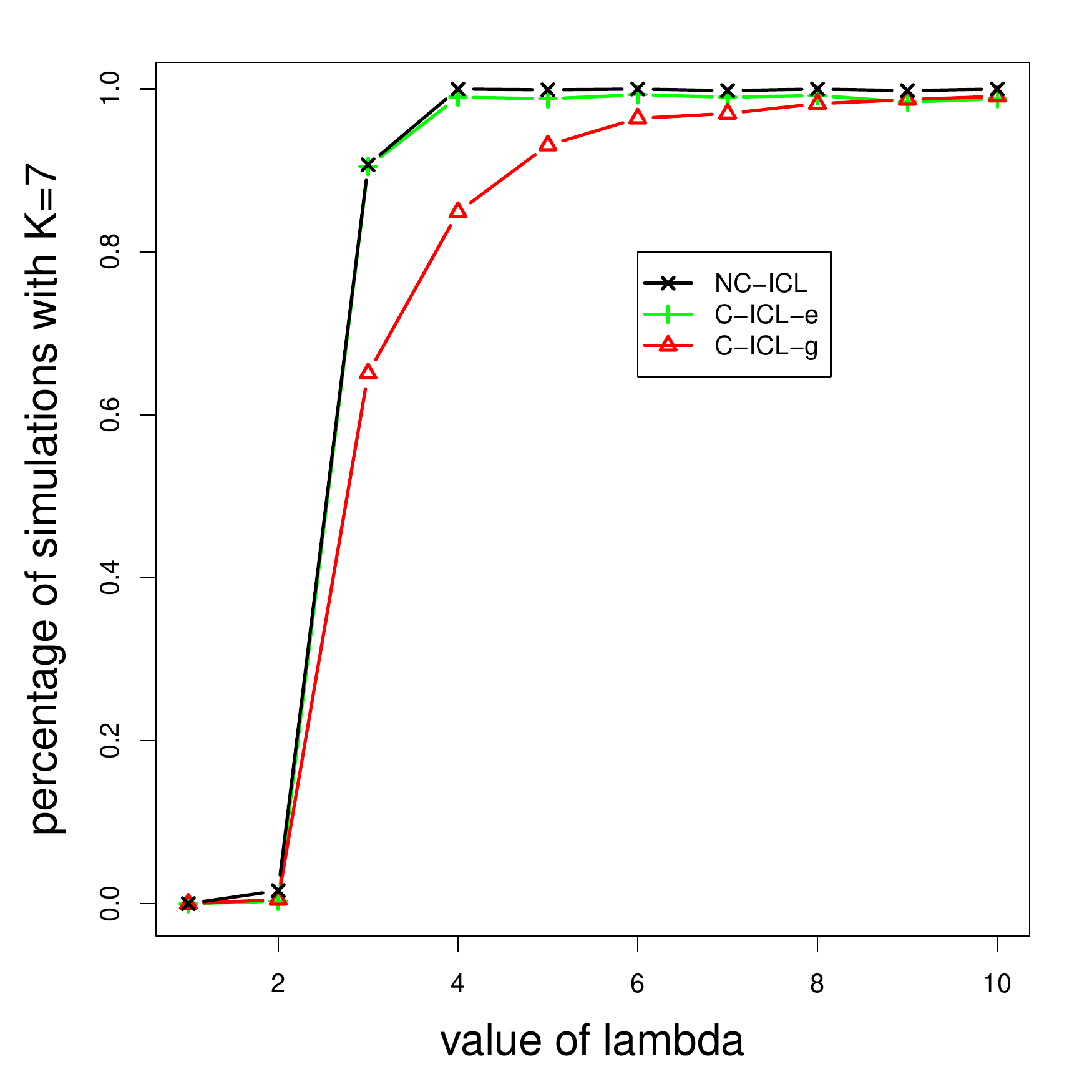} 
\end{center}
\caption{{\bf Performance of our method on small datasets.} Out of a thousand simulations on datasets of length $n=500$, percentage of times where each criterion, non-conditional ICL, conditional ICL with greedy initialization and conditional ICL with exact initialization, selected the appropriate number of segments, $K=7$, as the segment-level ratio increases. } \label{compa_small}
\end{figure}

Figure \ref{compa_small} summarizes the results of the simulation study for simulations of length 500.
While the {non-conditional} ICL criterion had the highest amount of correct estimates of number of segments $\hat{K}$, the faster ICL with pruned PDPA performed almost as well. Of note, the average run-times of the methods were {$4.2$ seconds for the {non-conditional} approach, $0.001$ and $0.12$ seconds respectively for the initialization of postCP with the k-means and PDPA algorithms, and $0.46$ seconds for the postCP algorithm.} 

\begin{figure}[!ht]
\begin{center}
\includegraphics[width=8cm,angle=0]{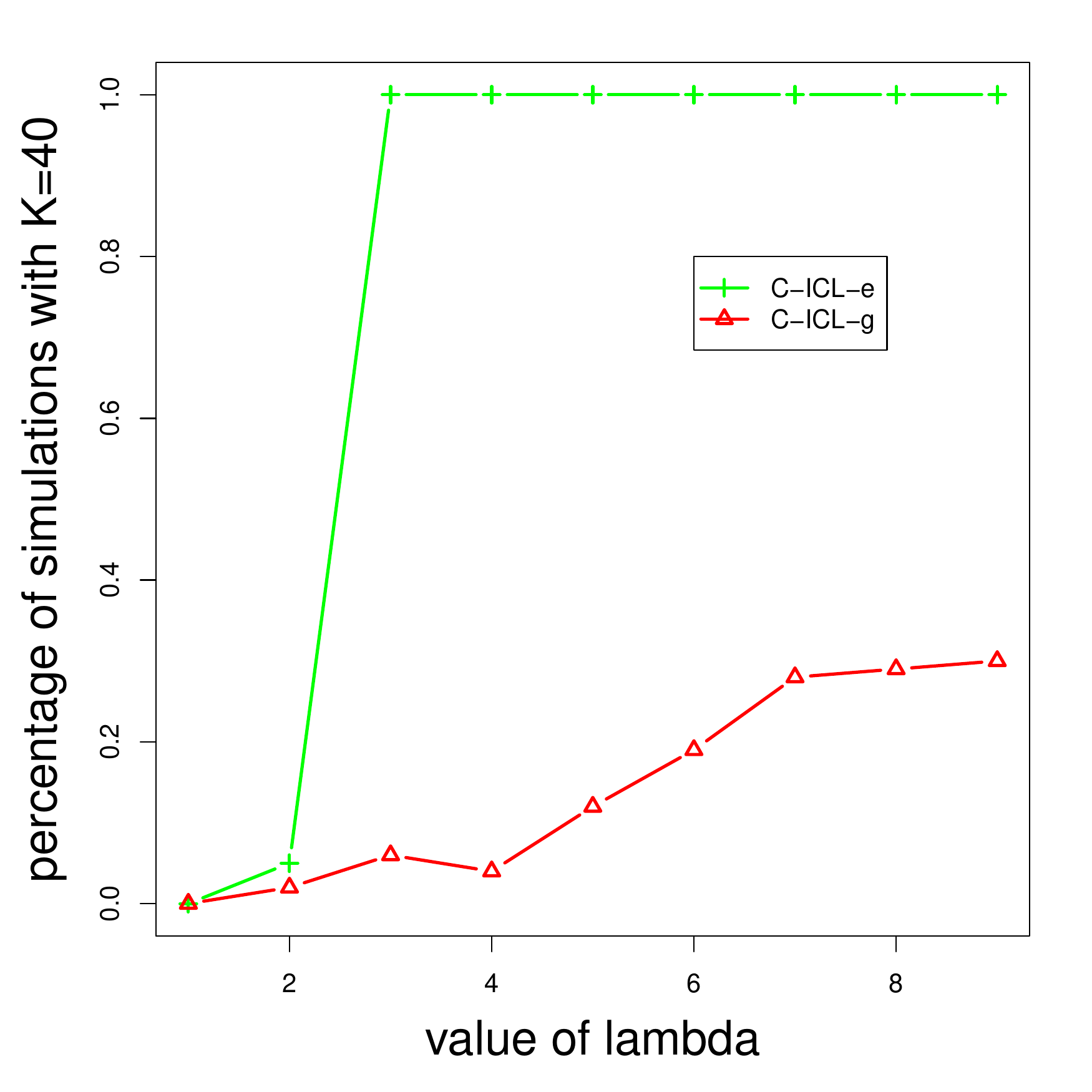} 
\end{center}
\caption{{\bf Performance of our method on large datasets.} Out of a hundred simulations on datasets of length $n=50,000$, number of times the conditional ICL criteria (with greedy and with exact initialization) selected the appropriate number of segments, $K=40$, as the segment-level ratio increases. }  \label{compa_long}
\end{figure}

Figure \ref{compa_long} summarizes the results of the simulation study for simulations of length $50,000$. For these larger sequences, the {conditional} ICL criteria performed much better when the initial change-point set was detected by PDPA than with the greedy algorithm. As the segmentation problem becomes more difficult with more segments, the greedy algorithm is less successful in providing accurate initial change-point location estimates. As a result, less accurate values of $\hat{\Theta_K}$ are used and {the conditional ICL} is not as effective in predicting the number of segments as in the smaller sample size example. 

On the other hand, {the conditional} ICL combined with PDPA detected the correct number of segments more than $80\%$ of the time with larger inter-segmental differences of $\lambda>2$.
The average run-time for the {initialization was $1.32$ seconds for k-means and $142$ seconds for PDPA, while the model selection procedure required on average $1,240$ seconds ($\approx20$ minutes).} Despite the longer run-time, it is advised to use the PDPA for model selection in very long sequences as it provides a more accurate set of change-points than greedy methods.

\section{Discussion} \label{analysis}

\subsection{Choice of $K_{\max}$}
Our strategy to compute the estimate of the ICL proceeds in two steps. First, we recover all the best segmentations in $1$ to $K_{max}$ segments. Then, using the parameters from all these $K_{\max}$ segmentations as an initialization, we run a forward-backward algorithm.

The initialization step takes on average an $\mathcal{O}(K_{\max}n\log n)$ complexity using the PDPA \citep[see][]{Pdpa}. The complexity of the second step is in $\mathcal{O}(K_{\max}n)$. Depending on the applications, it might be desirable or not to consider $K_{\max}$ of the order of $n$, \citep[see][for a discussion]{killick_pelt}. 
In the second case, our strategy is efficient. On the other hand, in the first case the initialization step is on average in $\mathcal{O}(n^2\log n)$ and at worst in $\mathcal{O}(n^3)$, while the second step is in $\mathcal{O}(n^2)$. The first step is thus the limiting factor. 

 When the goal is to initialize the HMM by recovering all the best segmentations of $1$ to $n$ segments, which we showed to be desirable for the quality of the procedure, there exists to our knowledge no faster algorithms to obtain an exact solution to this problem. Moreover, in any case, enumerating the $\sum_{k=1}^n k$ change-points of these $n$ segmentations is already quadratic in $n$.
An alternative is to use the binary segmentation heuristic \citep{venkatraman_faster_2007} which is on average in $\mathcal{O}( \log(K_{\max}n))$. In that case the limiting factor is the second step which still is quadratic in $n$.

Thus, we believe our strategy is most suited for the second case, when $K_{\max}$ is much smaller than $n$.
In the first case, when $K_{\max}$ is of the order of $n$, our strategy is at least quadratic in $n$ and its application is limited to medium size profiles.

\subsection{Re-sampling of yeast RNA-Seq data}

\begin{figure}[!ht]
\begin{center}
\subfigure[Re-sampling schema]{\includegraphics[scale=0.3,angle=0]{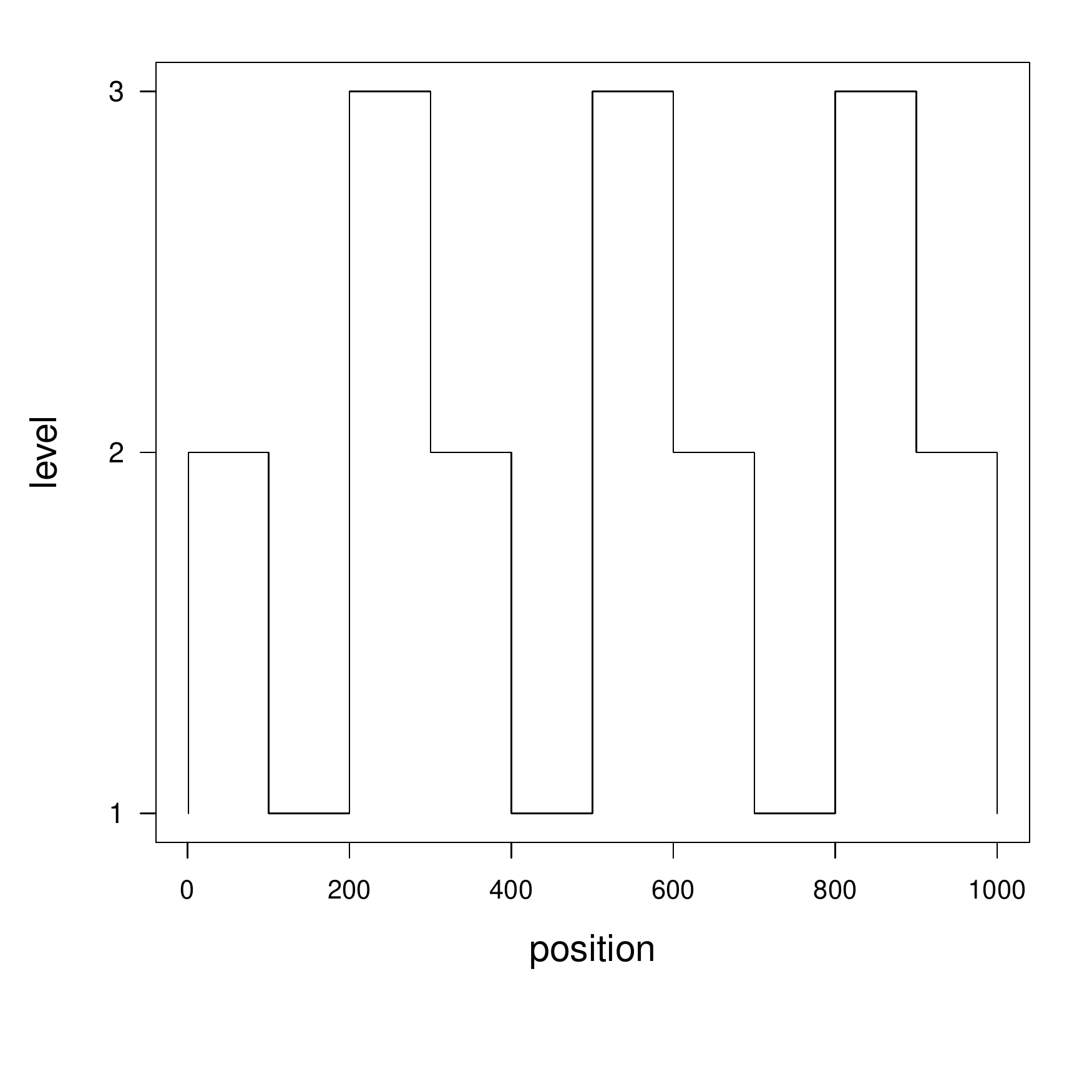}\label{plot1}}
\subfigure[Boxplot of number of segments $\hat{K}$]{\includegraphics[scale=0.3,angle=0]{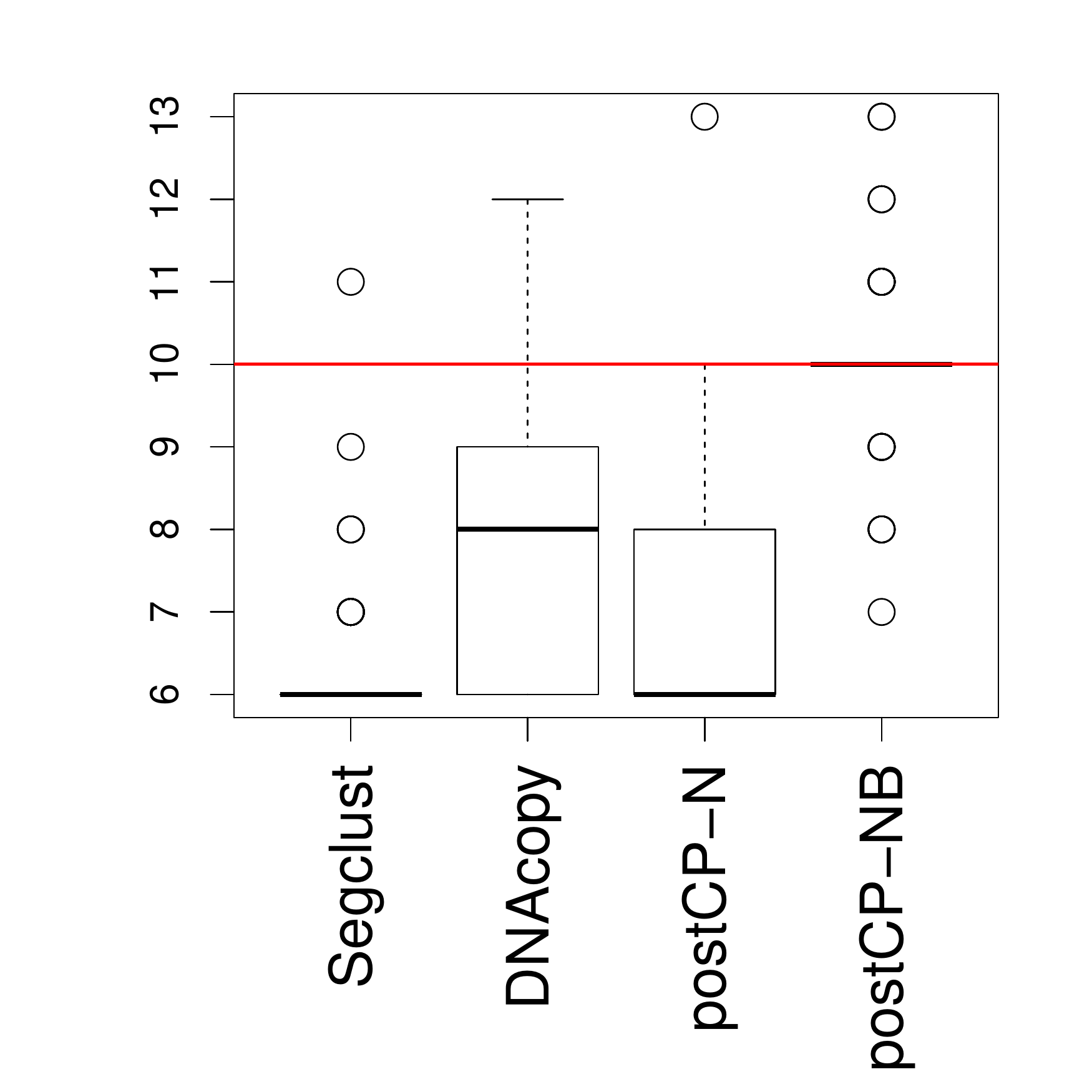}\label{box1}}
\end{center}
\caption{{\bf Algorithm comparison on short and regular dataset.} $n=1,000$ datapoints and $K=10$ equal segments. (a) Re-sampling schema displaying levels and lengths of segments. (b) Boxplot of estimated number of segments $\hat{K}$ for four different segmentation procedures for $100$ simulations.}
\end{figure}

\begin{figure}[!ht]
\begin{center}
\subfigure[Re-sampling schema]{\includegraphics[scale=0.3,angle=0]{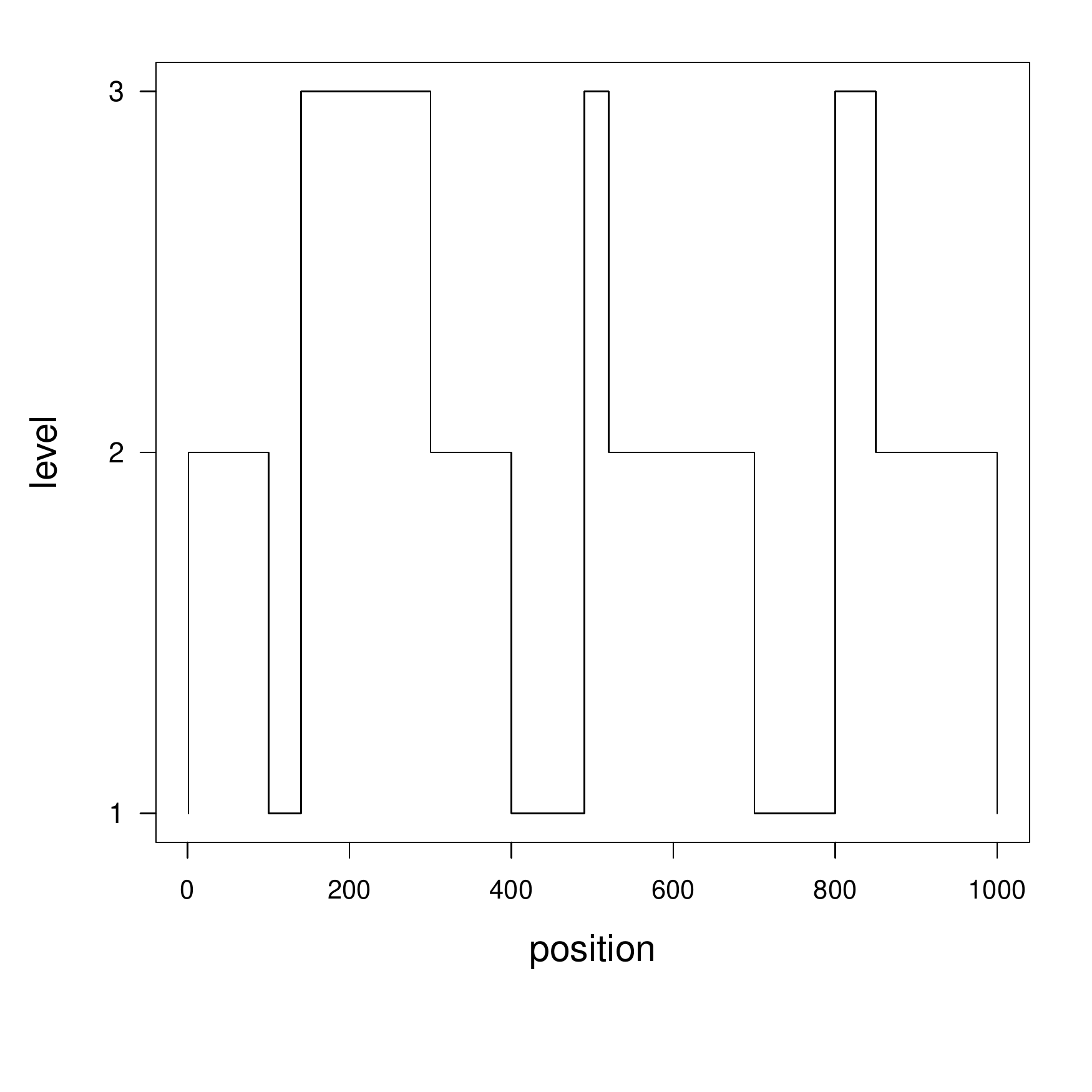}\label{plot2}}
\subfigure[Boxplot of number of segments $\hat{K}$]{\includegraphics[scale=0.3,angle=0]{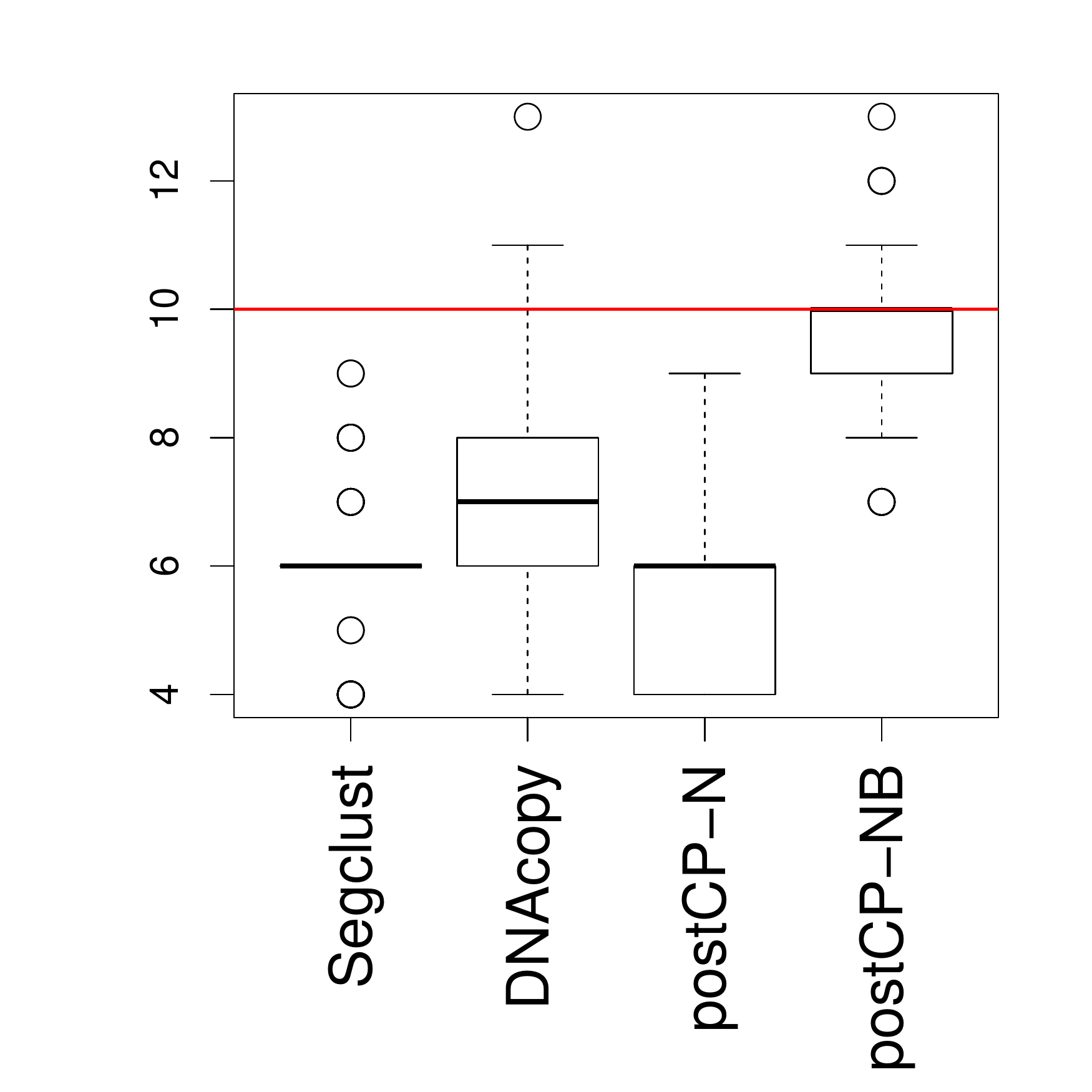}\label{box2}}
\end{center}
\caption{{\bf Algorithm comparison on short but irregular dataset.} $n=1,000$ datapoints and $K=10$ uneven segments. (a) Re-sampling schema displaying levels and lengths of segments. (b) Boxplot of estimated number of segments $\hat{K}$ for 4 different segmentation procedures for $100$ simulations.}
\end{figure}

\begin{figure}[!ht]
\begin{center}
\subfigure[Re-sampling schema]{\includegraphics[scale=0.3,angle=0]{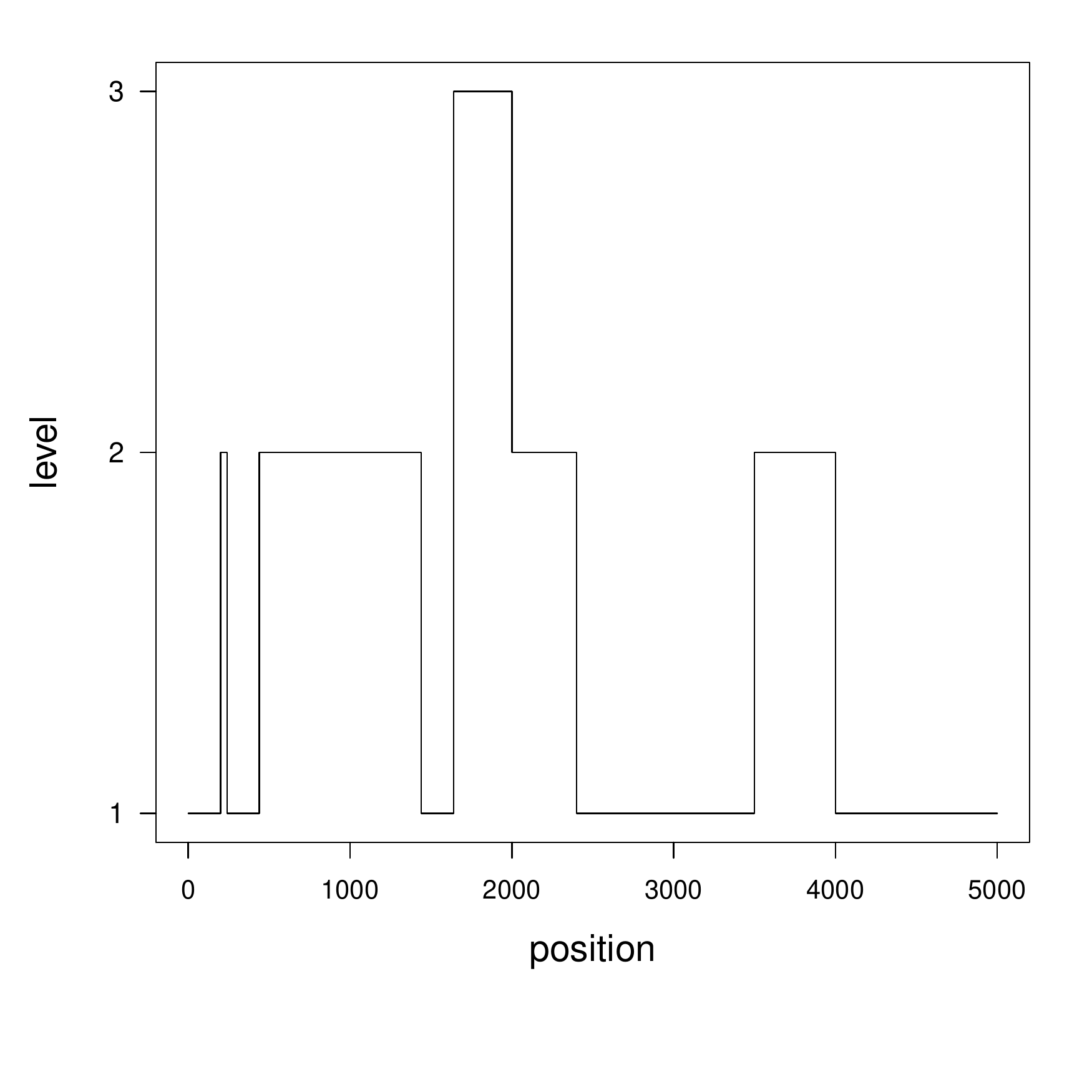}\label{plot3}}
\subfigure[Boxplot of number of segments $\hat{K}$]{\includegraphics[scale=0.3,angle=0]{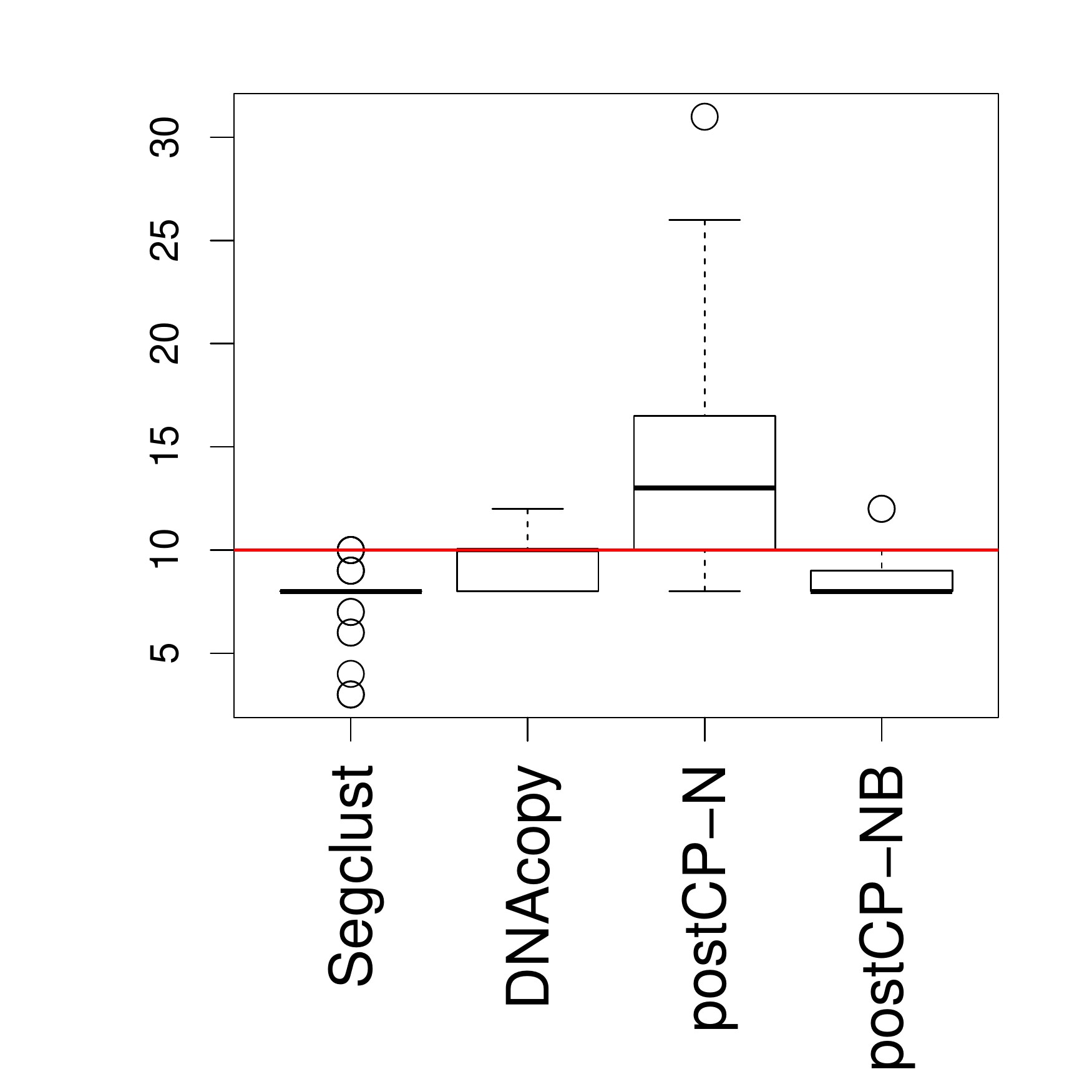}\label{box3}}
\end{center}
\caption{{\bf Algorithm comparison on medium length and irregular dataset.} $n=5,000$ datapoints and $K=10$ uneven segments. (a) Re-sampling schema displaying levels and lengths of segments. (b) Boxplot of estimated number of segments $\hat{K}$ for 4 different segmentation procedures for $100$ simulations.}
\end{figure}

\begin{figure}[!ht]
\begin{center}
\subfigure[Re-sampling schema]{\includegraphics[scale=0.3,angle=0]{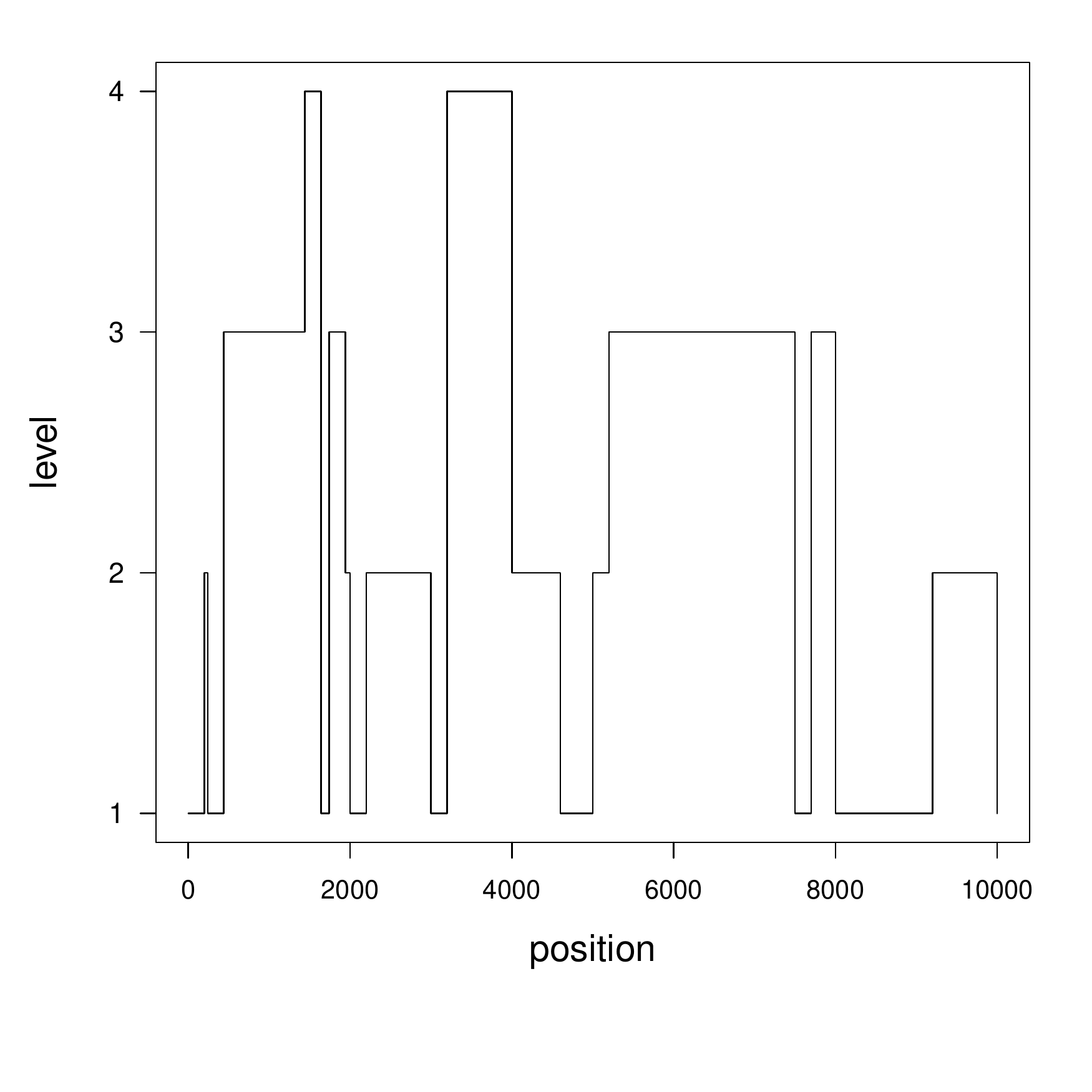}\label{plot4}}
\subfigure[Boxplot of number of segments $\hat{K}$]{\includegraphics[scale=0.3,angle=0]{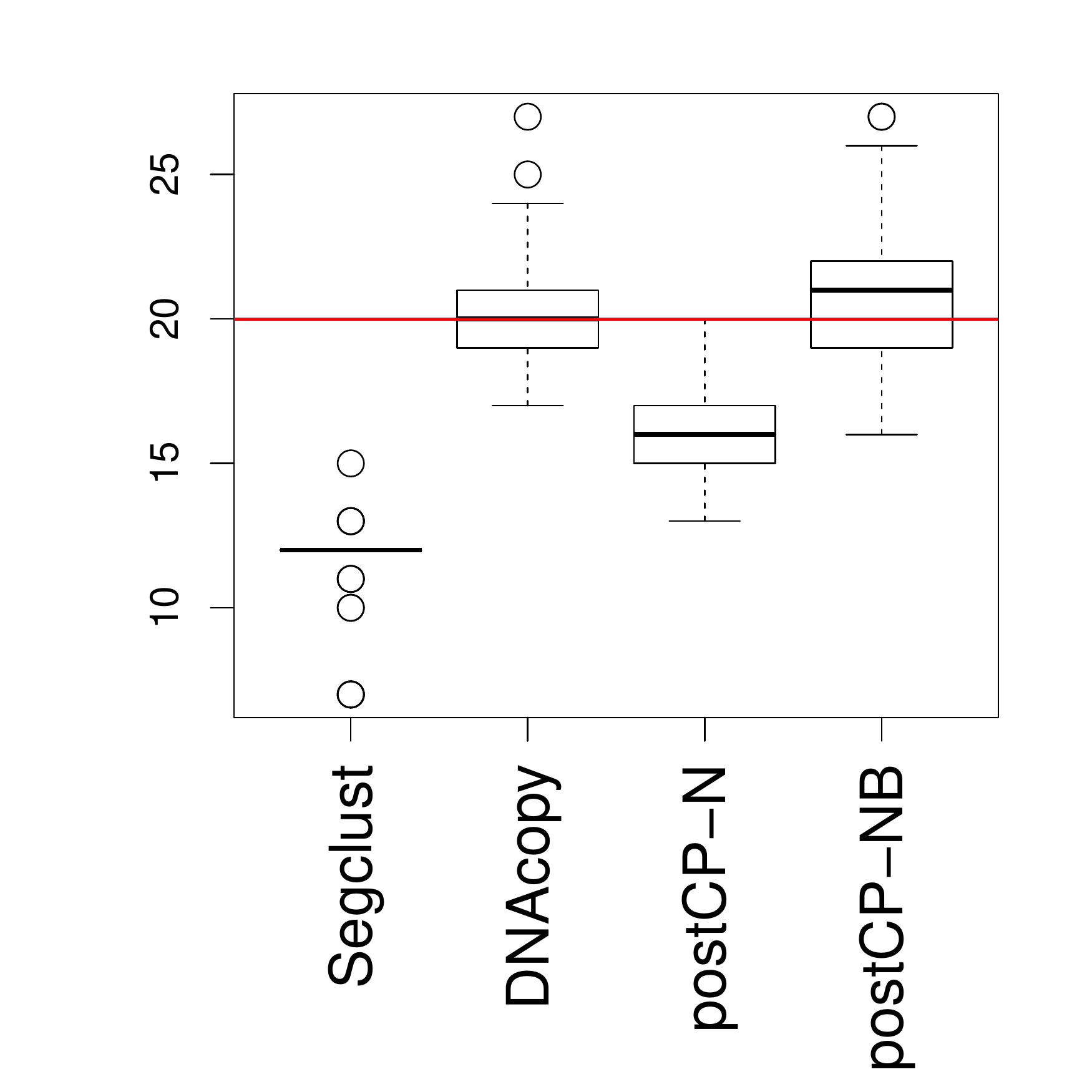}\label{box4}}
\end{center}
\caption{{\bf Algorithm comparison on long and irregular dataset.} $n=10,000$ datapoints and $K=20$ uneven segments. (a) Re-sampling schema displaying levels and lengths of segments. (b) Boxplot of estimated number of segments $\hat{K}$ for 4 different segmentation procedures for $100$ simulations.}
\end{figure}

To assess the quality of our criteria, we performed the following simulation study to compare two previously published packages on CRAN, segclust \citep{picard_segclust}, {which uses adaptive penalized likelihoods} and DNA copy, {an implementation of binary segmentation for multiple change-points} \citep{venkatraman_faster_2007}, with our model selection method with the {conditional} ICL criterion. We performed the following re-sampling procedure using real RNA-seq data. The original data, from a study by the Sherlock Genomics laboratory at Stanford University, is publicly available on the NCBI's Sequence Read Archive (SRA, \url{http://www.ncbi.nlm.nih.gov/sra}) with the accession number SRA048710. We clustered the observed signal into the following classes: intronic region, low expressed, medium expressed and highly expressed genes that we will refer to as levels $1$, $2$, $3$ and $4$. We then designed four simulation studies, each repeated $100$ times, by varying the number and level of segments as well as the signal and segment sizes, as described in Figures \ref{plot1} through \ref{plot4}. On each segment, the data was obtained by re-sampling (with replacement) the observations in the classified clusters.

To assess the performance of {our fast ICL} approach in segmentation, we used three different distributions as the emission distribution $g_\theta(\cdot)$ a normal distribution (postCP-N), a Poisson distribution (postCP-P) and negative binomial (postCP-NB) and used PDPA to obtain the initial set of parameters. {In all cases, we used homogeneous Markov chains with uniform priors; it is of note that the results of the constrained HMM methods may improve with informative priors \citep{fearnhead_exact_2005}, for example those taken from \textit{a posteriori} estimates}. For segclust, DNAcopy, and postCP-N, which assume a normal distribution, we applied the methods to the data after the widely used $\log(x+1)$ transformation. In all our simulation studies, postCP-P grossly overestimated the number of segments, so the results are not displayed here.\newline

In the simplest case, Figure~\ref{plot1} illustrates the resampling scheme for $n=1,000$ and $K=10$ evenly spaced segments, displaying the levels used for each segment and the change-point locations. Figure~\ref{box1} displays a boxplot of the number of segments found by each approach. In this quite unrealistic scenario, postCP-BN estimated the correct number of segments in 63 of 100 replicates. The next best algorithms were postCP-N and DNAcopy, respectively, which both slightly underestimated the number of segments. The segclust procedure provided a consistent underestimate of the number of segments.

Figures~\ref{plot2} and \ref{box2} illustrates the re-sampling schemes and boxplots for a slightly different and more realistic scenario of $n=1,000$ and $K=10$, with unevenly spaced segments this time. The results are comparable to the previous except that the methods performed slightly worse; the median postCP-NB estimate was still correct but missed 1 or 2 segments in 43 replicates. This suggests that postCP has more difficulties in detecting small segments.

We then replicated the methods for larger data sets and unevenly spaced segments. Figures~\ref{plot3} and \ref{box3} display the methods and results for a $n=5,000$ and $K=10$ scenario. In this case, DNAcopy performs best, with the median number of estimated segments being correct. The postCP-NB method gave similar results but missed two change-points in 66 of the replicates.
The segclust algorithm, once again, found consistent but overly conservative estimates of the number of segments, while postCP-N grossly overestimated the segments as the log-transformation was not adequate in this design. 

To understand the results, we ran the PDPA on the simulated datasets to obtain the optimal segmentations w.r.t. to negative binomial likelihood imposing $K=10$ segments. We found that in $48$ replicates out of $100$, this segmentation did not include the second true segment but rather sheared other segments into more pieces. This finding suggests that, at least in these $48$ replicates, precisely finding the position of the first two changes might be {prohibitively} difficult. 
Thus by selecting $K=8$ change-points rather than $10$, postCP-NB is coherent with the goal of the ICL {(i.e. selecting a set of segments such that we are confident in the position of these changes).}

In a $n=10,000$ and $K=20$ scenario with uneven segments (Figures~\ref{plot4} and \ref{box4}), DNAcopy was again best, with postCP-N and postCP-NB almost as effective, the former method slightly underestimating the number of segments and the latter approach slightly overestimating them.\newline

In the investigated scenarios, we found postCP, when the correct negative binomial distribution was specified, provided accurate and precise results when segments were evenly spaced, but provided slightly less accurate results in more realistic scenarios where segment lengths were uneven. The results with postCP-N and postCP-P suggest that the postCP approach may be susceptible to misspecification of the emission distribution when there are very small segments present (Figure~\ref{box3}). 
{Given the goal of the ICL this is to be expected. {Indeed, it is reasonable to have high uncertainty in the identification of small segments when the emission distribution is misspecified.}}

On the other hand, DNAcopy tended to underestimate segments in easier scenarios, where segments where even, but obtained more accurate results with more realistic uneven segments. The hybrid segmentation and clustering approach, segclust, generally was consistent but underestimated the number of segments.

\subsection{Application to a real data-set}

{We finally illustrate the procedure on the real data-set from the Sherlock study described above, whose underlying characteristics are unknown. The signal corresponds to the positive strand of chromosome $1$ from the yeast genome and has a length of $230,218$.} 

{We used a negative binomial model with global overdispersion parameters and initialized our procedure using the pruned dynamic programming algorithm (for a runtime of $25$ minutes). The postCP algorithm then required $4$ hours to analyze the profile, resulting in a choice of $79$ segments.}

\begin{figure}[!ht]
\begin{center}
\includegraphics[scale=0.45,angle=0]{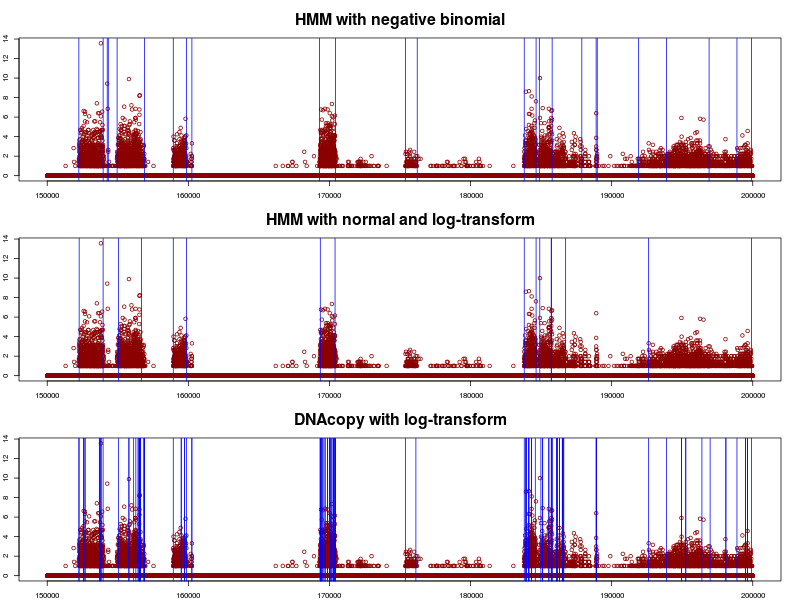}
\end{center}
\caption{{\bf Segmentation of yeast dataset.} The profile corresponding to the positive strand of chromosome 1 from the yeast genome is of length $230218$ and was segmented by three different methods. This figure illustrates the result on a region of the signal for our method with the negative binomial as emission distribution (Top), with Gaussian as emission (Middle) and for the DNAcopy algorithm.}\label{real}
\end{figure}

{We also compared these results to those proposed by the previously cited methods. However, we were not able to run the segclust algorithm on this long profile due to lack of memory capacity. With a similar runtime, the postCP algorithm with the normal distribution applied to the log-transformed data resulted in a choice of $80$ segments, while DNAcopy analyzed the signal in $47$ seconds for a final choice of $465$ segments. Figure \ref{real} illustrates the segmentation proposed by each method. For clarity, we focus on a region of length $50,000$ datapoints, and plotted the signal in a square-root scale. Even though the constrained HMM approach chooses almost the same number of segments with different emission distributions, their corresponding resulting segmentations differ.}

\subsection{Conclusion}\label{discussion}

We describe a fast procedure for estimating the ICL criterion in the context of model selection for segmentation. While simulations showed that the performance of the {conditional} $\ICL$ approach was almost as good as that of the {non-conditional} approach, several features allow for its use in a wide range of applications. The described ICL algorithm is versatile as it can be applied to data of any model distribution when provided with an initialization for the HMM, through either maximum likelihood estimation or the expectation-maximization (E-M) algorithm. 
{While there exists some model selection criteria that could be adapted to our problem such as the BIC or the MDL \citep{davis_2006} which provide a balance between data fitting and model complexity, the ICL also takes into account the entropy of the segmentation space. Given the very large collection of possible segmentations, we believe that the ICL is an interesting alternative to more standard model selection criteria.}

Furthermore, our procedure can be applied to long signals due to its fast run-time. With its effective results in finding the number of segments, specifically those where the precise location of the change-points can be estimated, this paper shows the practicality of the {conditional} $\ICL$ procedure in a wide variety of segmentation problems. 

\section*{Acknowledgments}

The authors would like to thank St\'ephane Robin for useful discussions.
Alice Cleynen's research was supported by an Allocation Special Normalien at the Universit\'e Paris-Sud in Orsay and The Minh Luong's research was supported by an excellence postdoctoral grant at the Universit\'e Paris-Descartes.





\bibliographystyle{plainnat}
\bibliography{Biblio}

\end{document}